\documentclass[10pt]{article}

\usepackage[top = 30truemm, bottom = 30truemm, left = 25truemm, right=25truemm]{geometry}
\usepackage{graphicx, color,amsmath}

\usepackage{indentfirst}
\begin{document}

\vspace*{2em}
\begin{center}
\LARGE\bf
Impacts of nuclear-physics uncertainties\\ in the s-process determined by Monte-Carlo variations
\end{center}
\vspace{1em}
{\large
N.~\textsc{Nishimura}$^{1}$\footnote{E-mail: nobuya.nishimura@yukawa.kyoto-u.ac.jp}\footnote{BRIDGCE UK Network; www.bridgce.ac.uk, UK}, G.~\textsc{Cescutti}$^{2}$\footnotemark[2], R.~\textsc{Hirschi}$^{3,4}$\footnotemark[2], T.~\textsc{Rauscher}$^{5,6}$\footnotemark[2], J.~\textsc{den~Hartogh}$^3$\footnotemark[2] and A. St. J. \textsc{Murphy}$^{7}$\footnotemark[2]}
\\
\\
\hspace*{1em}$^1$\,Yukawa Institute for Theoretical Physics, Kyoto University, Kyoto 606-8502, Japan\\
\hspace*{1em}$^2$\,INAF, Osservatorio Astronomico di Trieste, via G.B. Tiepolo 11, I-34131 Trieste, Italy\\
\hspace*{1em}$^3$\,Astrophysics Group, Keele University, Keele ST5 5BG, UK\\
\hspace*{1em}$^4$\,Kavli IPMU (WPI), University of Tokyo, Kashiwa 277-8583, Japan\\
\hspace*{1em}$^5$\,Department of Physics, University of Basel, 4056 Basel, Switzerland\\
\hspace*{1em}$^6$\,Centre for Astrophysical Research, University of Hertfordshire, Hatfield AL10 9AB, UK\\
\hspace*{1em}$^7$\,School of Physics and Astronomy, University of Edinburgh, Edinburgh EH9 3FD, UK\\

\begin{abstract}\noindent The s-process, a production mechanism based on slow-neutron capture during stellar evolution, is the origin of about half the elements heavier than iron. Abundance predictions for s-process nucleosynthesis depend strongly on the relevant neutron-capture and $\beta$-decay rates, as well as on the details of the stellar model being considered. Here, we have used a Monte-Carlo approach to evaluate the nuclear uncertainty in s-process nucleosynthesis. We considered the helium burning of massive stars for the weak s-process and low-mass asymptotic-giant-branch stars for the main s-process. Our calculations include a realistic and general prescription for the temperature dependent uncertainty for the reaction cross sections. We find that the adopted uncertainty for (${\rm n},\gamma$) rates, tens of per cent on average, effects the production of s-process nuclei along the line of $\beta$-stability, and that the uncertainties in $\beta$-decay from excited state contributions, has the strongest impact on branching points.
\end{abstract}

\section{Introduction}

Nucleosynthesis of heavy elements beyond the iron-group peak (mass number $A \sim 60$) is distinctly different from the production process of lighter elements. Neutron-capture is considered to be the primary production mechanism of heavier nuclei, up to $A\sim200$, facilitated by the neutron having no electric charge, and thus enabling penetration of substantial Coulomb barriers. Two different neutron-capture processes have been proposed \cite{1957RvMP...29..547B}, {\it i.e.}, the s- and r-process that are slow and rapid as compared to $\beta$-decay half-lives, respectively. The s-process occurs in stellar environments that feature lower neutron densities, while environments with higher neutron densities allow the faster rate of captures that leads to r-process nucleosynthesis.

The slow timescale of the s-process means that it occurs in stellar burning environments that evole over longer timescale. There are two astronomical sites and corresponding classes of the s-process (see a review~\cite{2011RvMP...83..157K} and references therein). The {\it main} s-process occurs in (i) thermal pulses of low-mass asymptotic-giant-branch (AGB) stars producing heavy nuclei up to Pb and Bi, while the {\it weak} s-process takes place in helium-core and carbon-shell burnings of massive stars and involves lighter nuclei up to $A \approx 90$. In both cases, the primary mechanism is to produce heavier elements due to the neutron capture and $\beta$-decay along stable isotopes from seed Fe nuclei over a long-term stellar evolution period. The neutron source reactions for the s-process are $\alpha$-capture to different nuclei, where $^{13}{\rm C}(\alpha,{\rm n})^{16}{\rm O}$ and $^{22}{\rm Ne}(\alpha,{\rm n})^{15}{\rm Mg}$ are main reactions for the main and weak s-processes, respectively. The impact of these key fusion reactions has been well studied~\cite{2011RvMP...83..157K}.

A major remaining issue is the effect of the uncertainties of the individual (n,$\gamma$) and $\beta$-decay rates on the final nucleosynthesis products. As there are many reactions involved in the s-process, the overall uncertainty is not as straightforward as for the cases of neutron source and poison reactions, for which key reactions are already well identified.  More systematic studies based on the Monte-Carlo (MC) and statistical analysis techniques \cite{2015JPhG...42c4007I, 2016MNRAS.463.4153R, 2018MNRAS.474.3133N} are necessary for such problems.

In the present paper, we investigate the impact of uncertainty caused by nuclear-physics on the production of s-process elements, using the MC-based nuclear-reaction network (see, \cite{2017MNRAS.469.1752N,Cescutti2018} for details). Adopting simplified stellar models that reproduce typical s-process nucleosynthesis patterns, we apply realistic temperature-dependent uncertainty of nuclear reaction and decay rates. We evaluate the uncertainty of nucleosynthesis yields and identify key reactions that have significant impact on the final s-process abundances.

\section{Methods}

For the nucleosynthesis calculations, we use simplified 1-D stellar evolution models with solar metallicity. We follow nucleosynthesis evolution along the temporal history of the temperature and density from the initial abundances. The thermal evolution is treated as the time evolution for a ``trajectory'' as a single fluid component. We adopt $25 M_\odot$ massive star evolution model \cite{2004A&A...425..649H, 2008IAUS..255..297H} and $2M_\odot$ AGB star model calculated by the MESA code \cite{2011ApJS..192....3P}. We have confirmed that these trajectories reproduce a typical abundance pattern for the main and weak s-process, respectively.

\begin{figure}[ht]
\centering
\includegraphics[height=0.24\hsize]{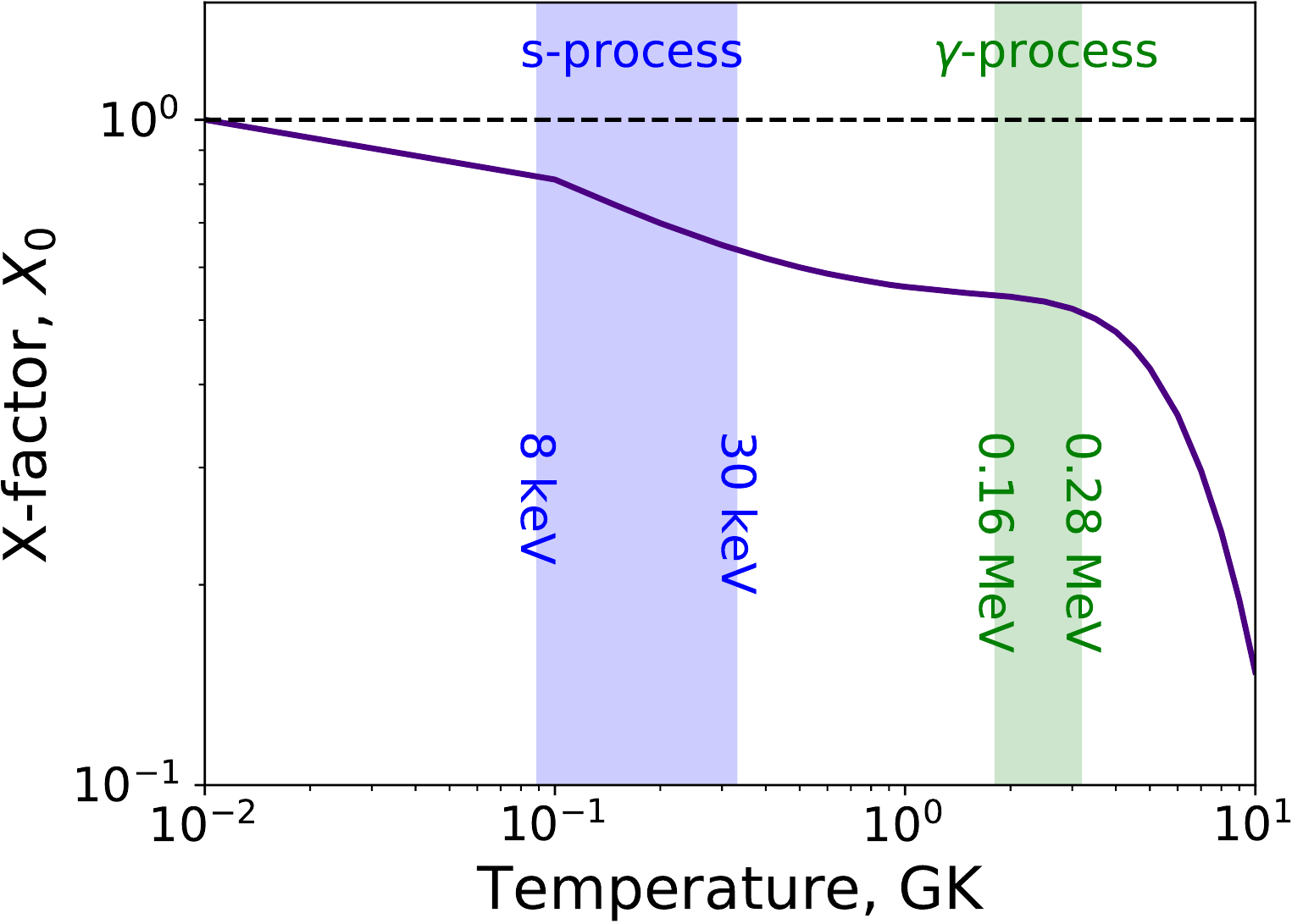}
\includegraphics[height=0.24\hsize]{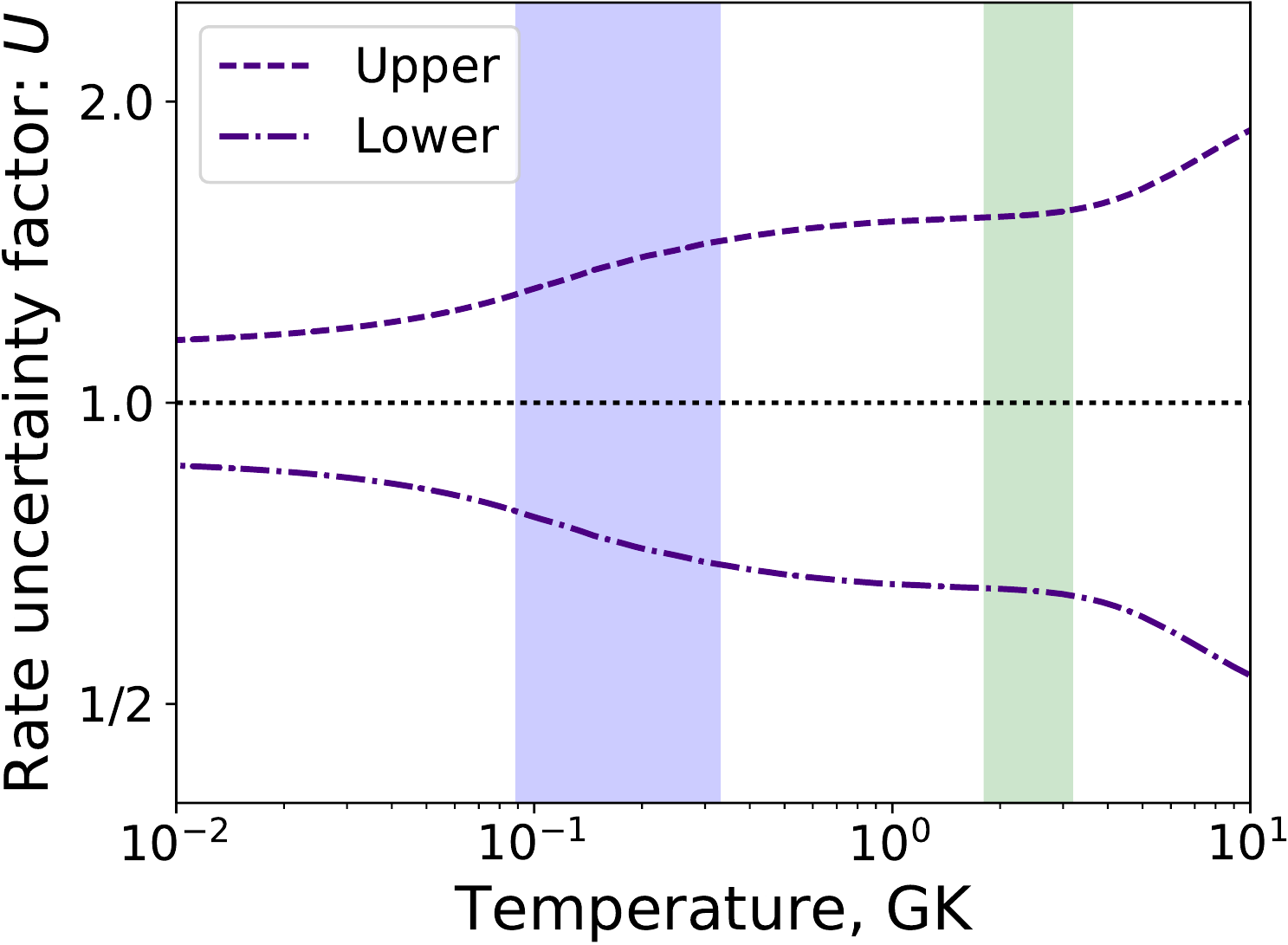}
\includegraphics[height=0.24\hsize]{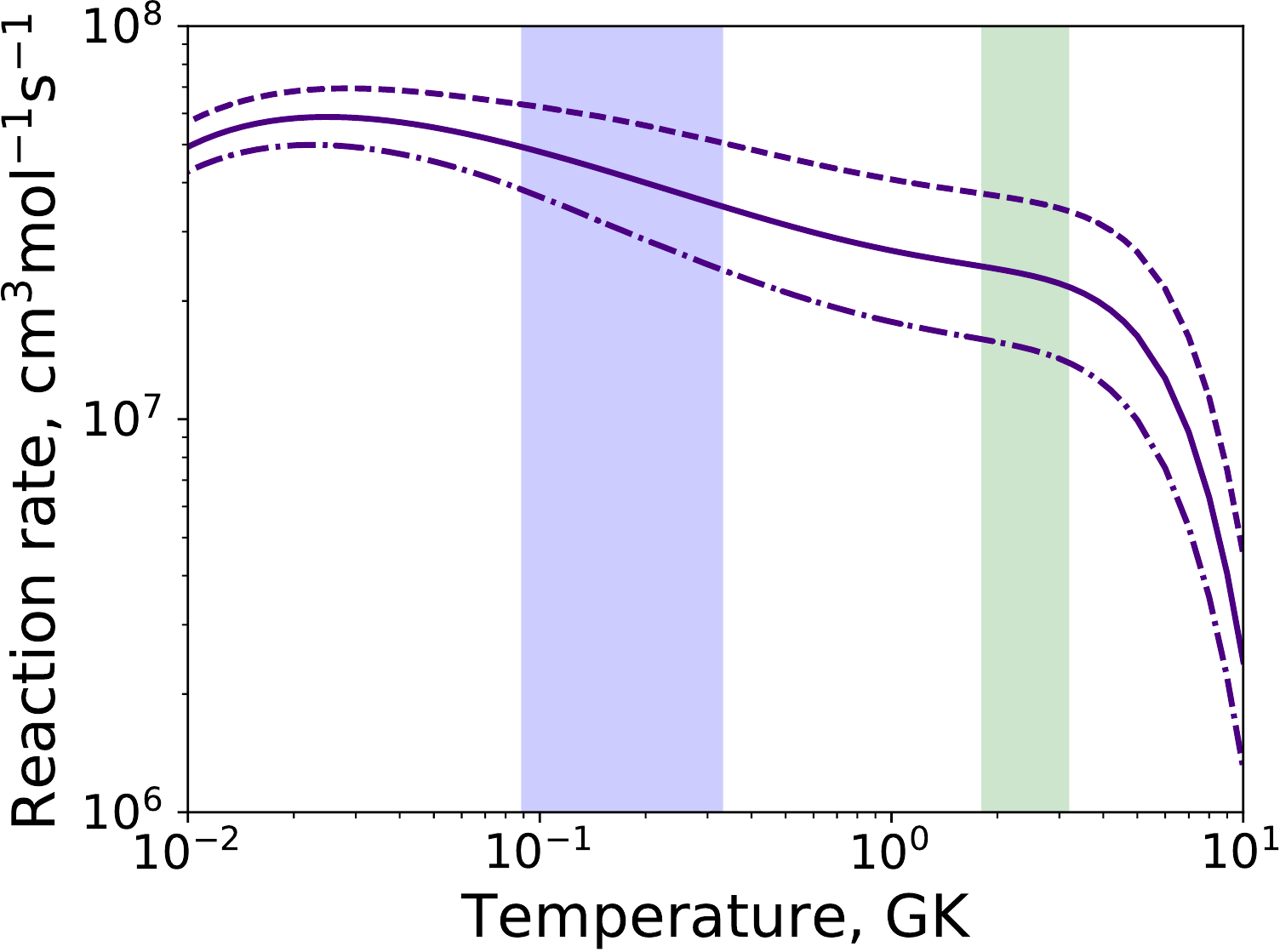}
\caption{Uncertainty factors for ${}^{83}{\rm Kr}({\rm n},\gamma){}^{84}{\rm Kr}$. ({\it Left}) $X_0$ adopted from ref~\cite{2011ApJ...738..143R}; ({\it middle}) the upper and lower limits of uncertainty factors; ({\it right}) the reaction rate \cite{2006AIPC..819..123D} with the upper and lower limits. In each panel, the temperature regions for the s-process (blue) and the $\gamma$-process (green) are highlighted.}
\label{fig-rate-unc}
\end{figure}

Many nuclear reaction rates of neutron capture relevant to the s-process have been experimentally measured, because the target nuclei are stable (this is not the case for other heavy-element nucleosynthesis, {\it e.g.}, the r-process, as reactions on the unstable nuclei play a major role). However, these experimental measurements are unable to measure the relevant $({\rm n},\gamma)$ reaction rates as realized at the high temperatures of the stellar environment due to contributions of excited states \cite{2011ApJ...738..143R, 2012ApJS..201...26R}. Therefore, we consider that reaction rates have a temperature-dependent uncertainty due to the relative contributions by the ground state and excited states for experimental based cross sections. Following the prescription in refs.~\cite{2011ApJ...738..143R, 2012ApJS..201...26R}, we apply the uncertainty factor $u(T)$ for thermonuclear reaction rates as
\begin{equation}
	\label{eq-unc}
	u(T) = X_0(T) u_{\rm exp} + \left[1 - X_0(T)\right]u_{\rm th}
\end{equation}
where $X_0$ is the temperature dependence factor and $u_{\rm exp}$ and $u_{\rm th}$ are uncertainty ranges for experimental and theoretical contributions, respectively. The value of $X_0(T)$ for ${}^{83}{\rm Kr}({\rm n},\gamma){}^{84}{\rm Kr}$ is shown in Figure~\ref{fig-rate-unc} (left panel), which decreases as the temperature increases from $1$ at lower temperatures (below $\sim 0.01$~GK). From Equation~\ref{eq-unc}, therefore, $u(T) \sim u_{\rm exp}$ at lower $T$, while $u(T)$ reaches $u_{\rm th}$ at higher $T$.

In this study, experimental uncertainties are used for the ground state contributions to (n,$\gamma$) rates, whereas a factor $2$ is used for excited state uncertainties (for details, see \cite{2012ApJS..201...26R, 2017MNRAS.469.1752N, 2018MNRAS.474.3133N}). As theoretical calculated rates may have large uncertainty, we simply apply a constant value $2$. We apply $u(T)$ to determine the upper limit and lower limit for the variation of reaction rates by multiplying $u(T)$ and $1/u(T)$, respectively. The middle panel of Figure~\ref{fig-rate-unc} shows the adopted uncertainty factor, while the right panel shows the uncertainty range for the ${}^{83}{\rm Kr}({\rm n},\gamma){}^{84}{\rm Kr}$ reaction.

A similar approach is used for $\beta$-decay rates, based on temperature-dependent partition functions $G(T)$ to determine the importance of excited states, {\it i.e.}, the uncertainty factor of $\beta$-decay rates $u_{\rm weak}$ is defined as
\begin{equation}
	u_{\rm weak} = \frac{2J_0 + 1}{G(T)} u^{\rm weak}_{\rm exp} 
					+ \left(1 - \frac{2J_0 + 1}{G(T)}\right) u^{\rm weak}_{\rm th} \ ,
\end{equation}
where $u_{\rm exp}$ and $u^{\rm weak}_{\rm th}$ are experimental and theoretical uncertainty factors, respectively. The uncertainty at lower temperatures ($T < 10^7$~K) corresponds to the measured value at the ground state ($u^{\rm weak}_{\rm exp}$), while the uncertainty becomes larger as the temperature increases. We adopt $u^{\rm weak}_{\rm th} = 1.3$ and $u^{\rm weak}_{\rm th} = 10$, of which the total uncertainty reaches up to $\sim 2$ in stellar burning temperatures.

\section{Results}

\subsection{Uncertainties of the s-process}

Our MC performs many nucleosynthesis simulations, each of which has each nuclear reaction rate sampled from an underlying distribution ({\it i.e.} applying the variation factor. A uniform random distribution between the upper and lower limit of the reaction rate at a given temperature was used for each variation factor. To identify the separate contributions from uncertainties in (n,$\gamma$) and from $\beta$-decay rates, we have performed three different cases: {\tt ngbt}, in which all (n,$\gamma$) and $\beta$-decay rates are varied; {\tt ng} where only (n,$\gamma$) rates are varied; and {\tt bt} in which only $\beta$-decay rates vary.

\begin{figure}[ht]
\centering
\includegraphics[height=0.328\hsize]{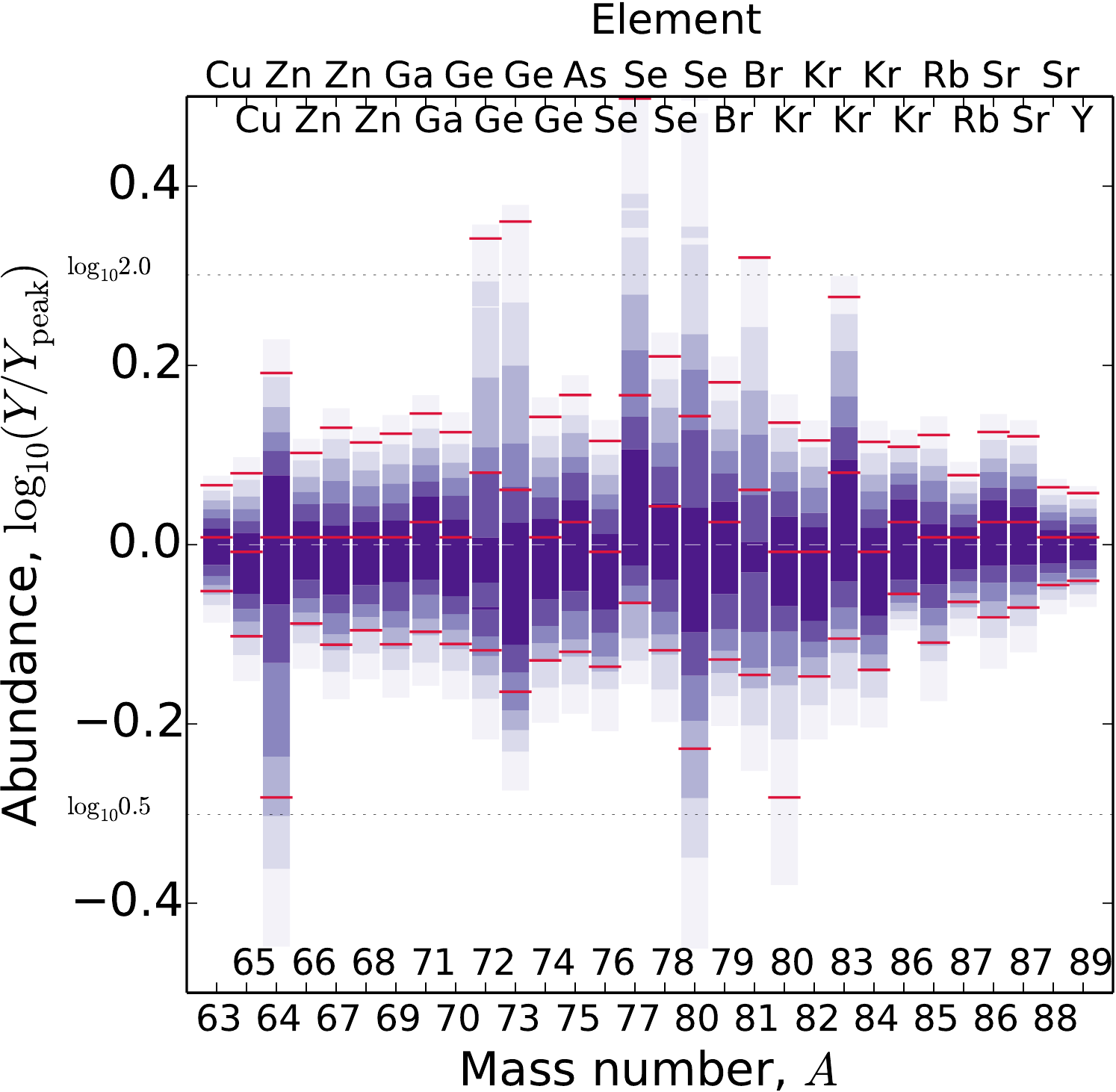}
\includegraphics[height=0.328\hsize]{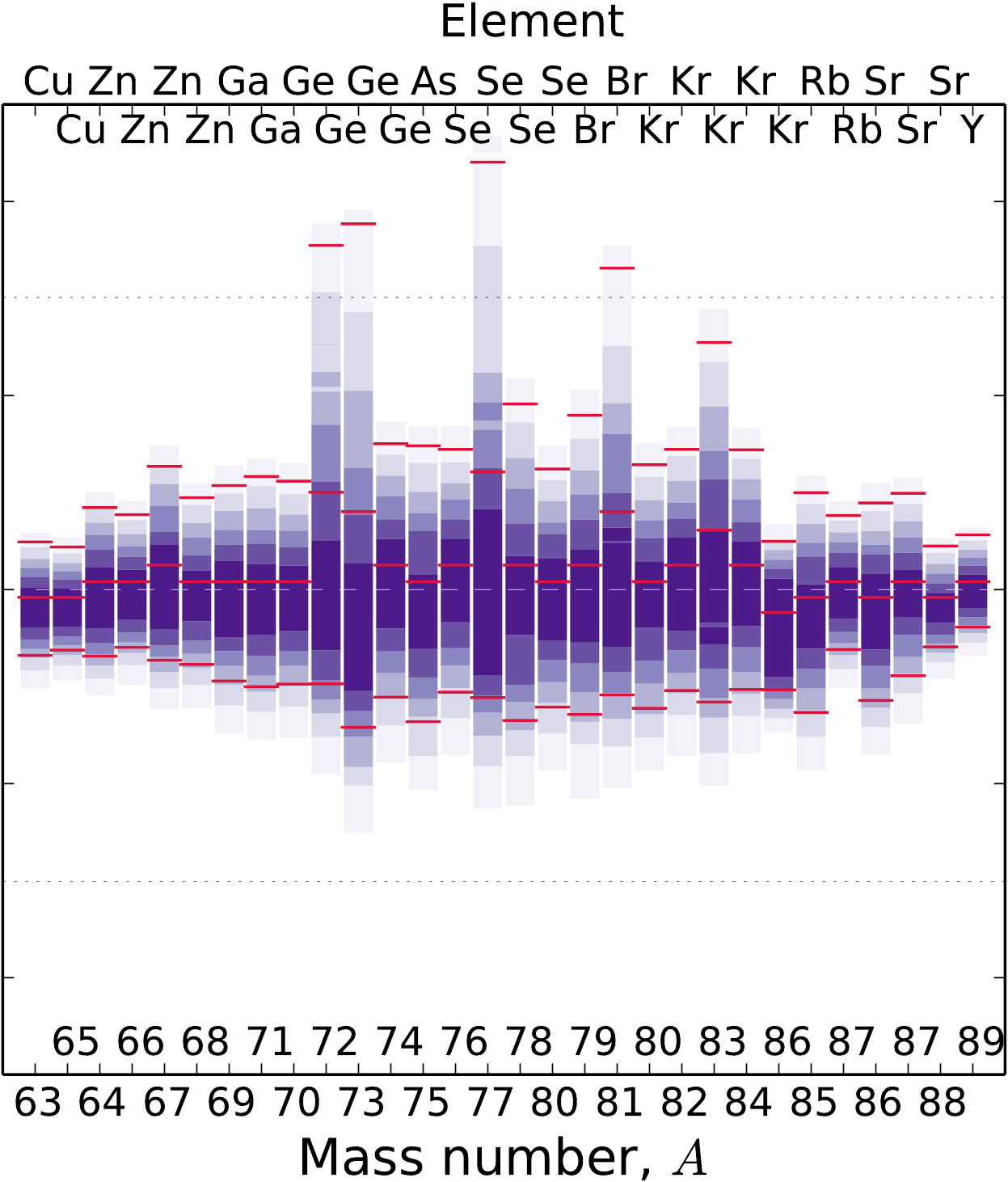}
\includegraphics[height=0.328\hsize]{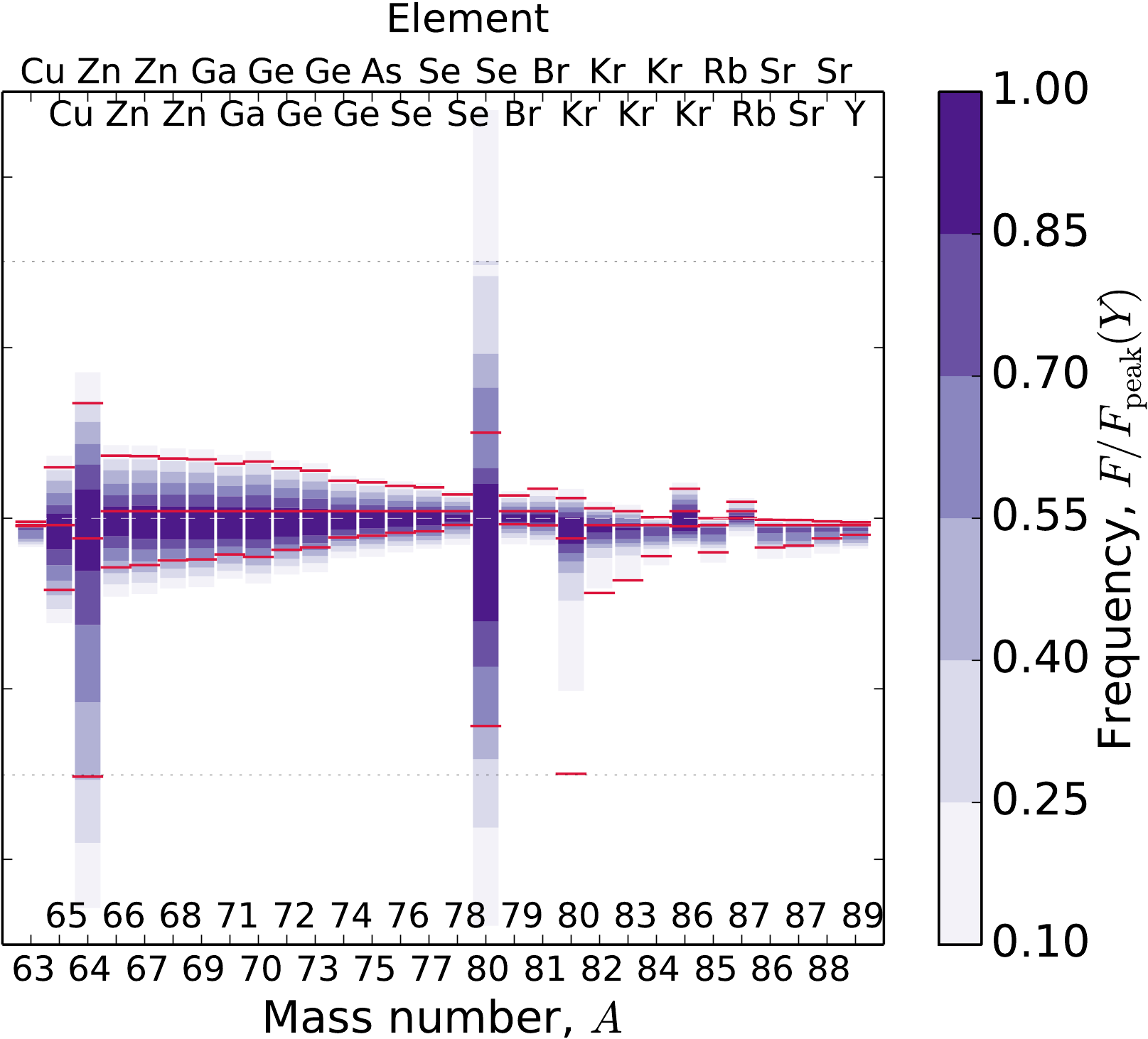}
\caption{The results of the MC for the weak s-process. The uncertainty range is shown for each isotope with red  lines covering $90$\% from the peak value for variation models of {\tt ngbt} ({\it left}), {\tt ng} ({\it middle}) and {\tt bt} ({\it right}).}
\label{fig-mc-ws}
\end{figure}

\begin{figure}[h]
\centering
\includegraphics[width=\hsize]{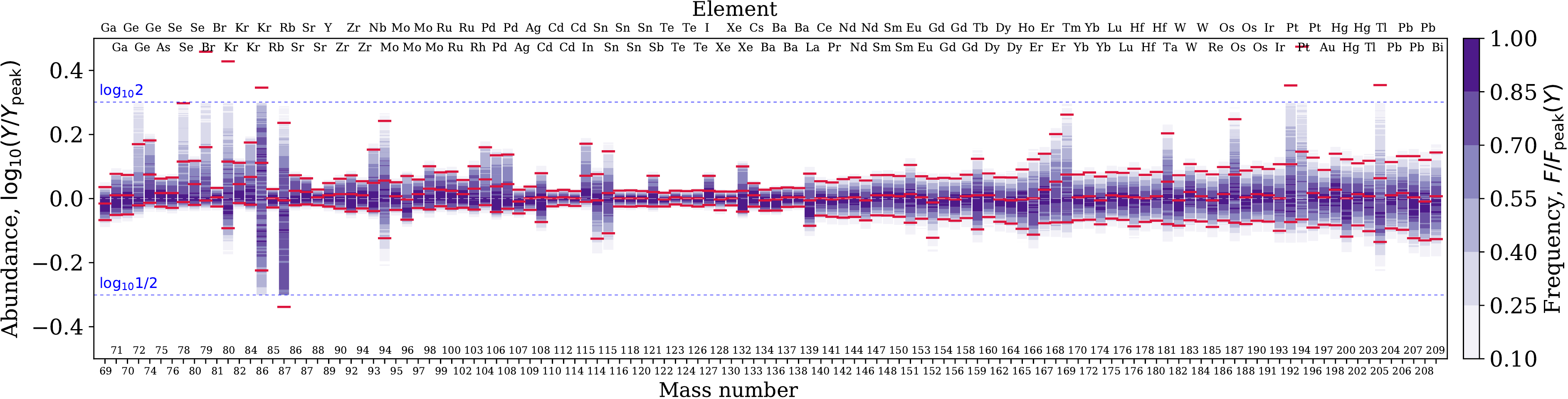}
\caption{The same as Fig.~\ref{fig-mc-ws}, but the results of main s-process.}
\label{fig-mc-ms}
\end{figure}

Fig.~\ref{fig-mc-ws} shows the resulting production uncertainty for the weak s-process for the cases where we varied all (n,$\gamma$) reactions and $\beta$-decays. We select abundance uncertainties for stable s-process isotopes up to $\sim 90$. The colour distribution corresponds to the normalized probability density distribution of the uncertainty in the final abundance.

Conidering the {\tt ngbt} case, the $90\%$ uncertainty range of abundances for most nuclides is less that a factor of $1.5$ ($0.176$ in $\log_{10}$) region, while some isotopes show a larger uncertainty that reaches factor $2$. Comparison of {\tt ng} and  {\tt bt} cases reveals that this is mostly due to (n,$\gamma$) reaction. Uncertainties for a few isotopes ($^{64}{\rm Zn}$ and $^{80}{\rm Se}$) are affected by $\beta$-decay around branching points, although the effects of $\beta$-decay to the global isotopes are minor compared with (n,$\gamma$). 

The impacts of $\beta$-decay uncertainties on the s-process appear only around s-process branchings. This is seen in the results of {\tt bt} (in Fig.~\ref{fig-mc-ws} and \ref{fig-mc-ms}), where a few $\beta$-decays cause larger uncertainties in nucleosynthesis. Our technique allows one to quantitatively analyze the MC result to identify the correlation between decay rate and final abundance (see, \cite{2017MNRAS.469.1752N}). We find that $^{64}{\rm Cu} (\beta^{+}) ^{64}{\rm Zn}$ and $^{80}{\rm Br} (\beta^{+}) ^{80}{\rm Kr}$ have the dominant impact on the production of $^{64}{\rm Zn}$ and $^{80}{\rm Se}$ for the weak s-process, respectively. These $\beta$-decay rates are around the s-process branching points as indicated in previous investigations (in Fig.~\ref{fig-mc-ws}).

These features are also pronounced for the case of main s-process, as the primaly physical mechanism is the same as in the weak s-process. The overall uncertainty of final abundances, shown in Figure~\ref{fig-mc-ms}, shows that they mostly caused by uncertainty of (n,$\gamma$) reactions except at branching points (see \cite{Cescutti2018} for more details). The impacts of $\beta$-decay uncertainties on the s-process appear only around s-process branchings. 

\subsection{Key neutron-capture reactions}

\begin{figure}[ht]
\centering
\includegraphics[width=\hsize]{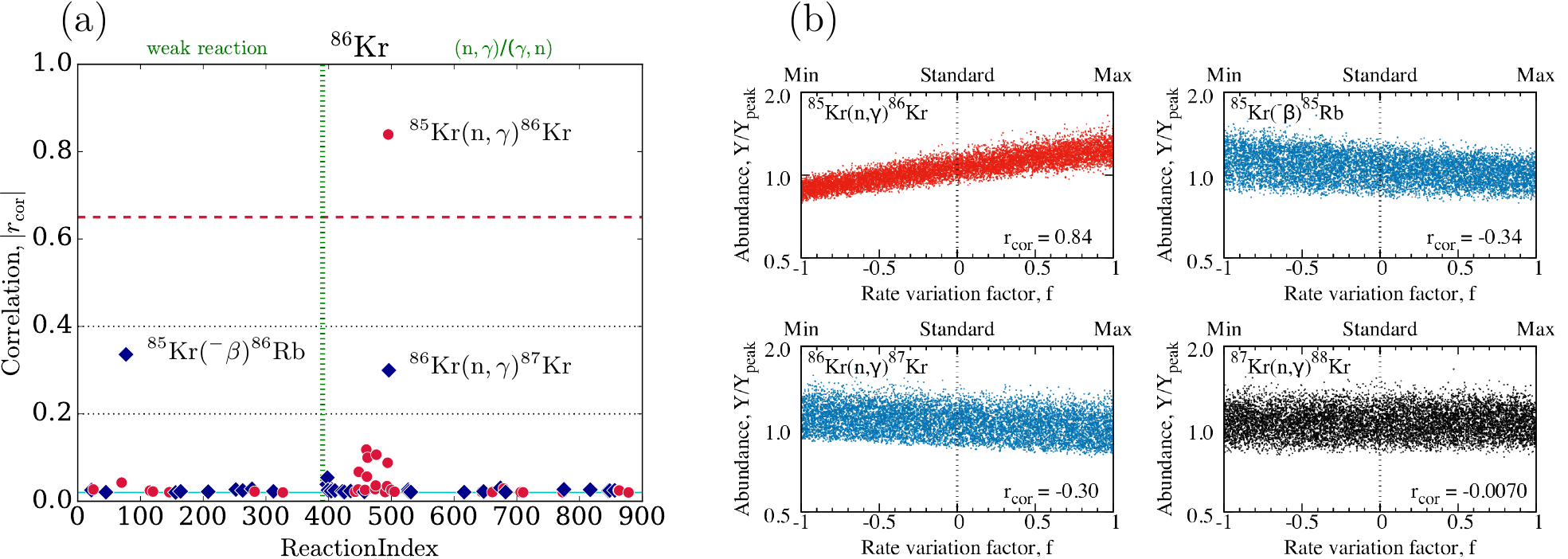}
\caption{The results of the MC for the weak s-process, focusing on $^{85}{\rm Kr}$ production. (a) The values of correlation factor, $|r_{\rm cor}|$, for all varied rates. (b) The distribution of uncertainty factor vs $|r_{\rm cor}|$ for selected reactions:.}
\label{fig-cor}
\end{figure}

Based on the MC calculations, we selected reactions \& decays that had a significant impact on the final abundance uncertainties. As shown in Fig.~\ref{fig-cor}, we calculated the Pearson's product-moment correlation coefficient, $r_{\rm cor}$, between variation factors and calculated abundances. In this study, we assume $|r_{\rm cor}| \geq 0.65$ as the significant value. Thus, ${}^{85}{\rm Kr}({\rm n},\gamma){}^{86}{\rm Kr}$ is the key reaction for the production of ${}^{86}{\rm Kr}$ as shown in Fig.~\ref{fig-cor}. We also find two cases ({\it i.e.}, ${}^{85}{\rm Kr}({}^{-}\beta){}^{85}{\rm Rb}$ and ${}^{86}{\rm Kr}({\rm n},\gamma){}^{86}{\rm Kr}$) with $|r_{\rm cor}| > 0.2$, which are possibly key reactions if the uncertainty involving ${}^{85}{\rm Kr}({\rm n},\gamma){}^{86}{\rm Kr}$ were to become significantly reduced due to future work. They are actually the key reactions for ${}^{86}{\rm Kr}$ when we perform the MC run omitting the uncertainty of ${}^{85}{\rm Kr}({\rm n},\gamma){}^{86}{\rm Kr}$, as shown in \cite{2017MNRAS.469.1752N}. Fig.~\ref{fig-cor}(b) presents the distribution of uncertainty factors and obtained abundances. This confirms the basic features of $r_{\rm cor}$ that a positive $r_{\rm cor}$ results in a positive correlation of the two parameters, and vice versa, and that a larger $|r_{\rm cor}|$ results in a stronger correlation.

We calculated correlation factors for the all possible combinations of varied reaction rate and s-process products. As the full lists of these key rates are summarized in our papers \cite{2017MNRAS.469.1752N, Cescutti2018}, here, we only highlight ``Level~1'' key rates with highest priorities for the weak and main s-processes. In Table~\ref{tab-key-ws} and \ref{tab-key-ms}, we list key $({\rm n},\gamma)$ reactions, of which $|r_{\rm cor}| >= 0.65$, for the weak s-process and main s-process, respectively. Here, only the target nucleus is listed for $({\rm n},\gamma)$ reactions, e.g. the $({\rm n},\gamma)$-target nucleus ``${}^{67}{\rm Zn}$'' indicates ${}^{67}{\rm Zn}({\rm n},\gamma){}^{68}{\rm Zn}$.

We note that there are a few cases that the key nucleus is not the target-nucleus of the key $({\rm n},\gamma)$ reaction. This is due to the propagation of large uncertainties from ``upstream'' to ``downstream'' through the s-process nucleosynthesis flow. Besides the reactions listed in the tables, there are still some reactions that show the non-negligible value of $r_{\rm cor}$. These will become important reactions if the relevant first-priority reactions become well determined ({\it e.g.} through future experimental work). Refs \cite{2017MNRAS.469.1752N, Cescutti2018} provide further details.

\begin{table}[ht]
\centering
\caption{The key neutron-capture reactions for the weak s-process. Key $({\rm n},\gamma)$ reactions are listed with their correlation factors $r_{\rm cor,0}$ for each key ``product'' nucleus. Only the target nucleus for the involving neutron capture is shown in the column of ``$({\rm n},\gamma)$-target''.}
  \begin{tabular}{c|cccccccccccc}
  \hline
Product &  ${}^{67}{\rm{Zn}}$ &  ${}^{72}{\rm{Ge}}$ &  ${}^{73}{\rm{Ge}}$ &  ${}^{77}{\rm{Se}}$ &  ${}^{78}{\rm{Se}}$ &  ${}^{81}{\rm{Kr}}$ &  ${}^{83}{\rm{Kr}}$ &  ${}^{85}{\rm{Kr}}$ &\\
$({\rm n},\gamma)$-target &   ${}^{67}{\rm Zn}$ &   ${}^{72}{\rm Ge}$ &   ${}^{73}{\rm Ge}$ &   ${}^{77}{\rm Se}$ &   ${}^{78}{\rm Se}$ &   ${}^{81}{\rm Br}$ &   ${}^{83}{\rm Kr}$ &   ${}^{86}{\rm Kr}$ &\\
$r_{{\rm cor},0}$ &$ -0.67$ &$ -0.85$ &$ -0.84$ &$ -0.86$ &$ -0.71$ &$ -0.80$ &$ -0.76$ &$  0.84$ &\\
\hline
  \end{tabular}
  \label{tab-key-ws}
\end{table}

\begin{table}[ht]
\centering
\small

\caption{Key neutron-capture reactions for the main s-process. The columns are the same as Table~\ref{tab-key-ws}.}
\scalebox{0.92}{
  \begin{tabular}{c|ccccccccccccccc}
  \hline
Product &  ${}^{69}{\rm{Ga}}$ &  ${}^{71}{\rm{Ga}}$ &  ${}^{70}{\rm{Ge}}$ &  ${}^{72}{\rm{Ge}}$ &  ${}^{74}{\rm{Ge}}$ &  ${}^{75}{\rm{As}}$ &  ${}^{76}{\rm{Se}}$ &  ${}^{78}{\rm{Se}}$ &  ${}^{79}{\rm{Se}}$ &  ${}^{79}{\rm{Se}}$ &  ${}^{80}{\rm{Se}}$ &  ${}^{81}{\rm{Br}}$ &\\
$({\rm n},\gamma)$-target &   ${}^{69}{\rm Ga}$ &   ${}^{71}{\rm Ga}$ &   ${}^{70}{\rm Ge}$ &   ${}^{72}{\rm Ge}$ &   ${}^{74}{\rm Ge}$ &   ${}^{75}{\rm As}$ &   ${}^{76}{\rm Se}$ &   ${}^{78}{\rm Se}$ &   ${}^{79}{\rm Br}$ &   ${}^{80}{\rm Kr}$ &   ${}^{80}{\rm Se}$ &   ${}^{81}{\rm Br}$ &\\
$r_{\rm cor,0}$ &$ -0.78$ &$ -0.89$ &$ -0.87$ &$ -0.93$ &$ -0.97$ &$ -0.86$ &$ -0.89$ &$ -0.97$ &$ -0.94$ &$ -0.90$ &$ -0.96$ &$ -0.74$ &\\
\hline
Product &  ${}^{84}{\rm{Kr}}$ &  ${}^{85}{\rm{Kr}}$ &  ${}^{85}{\rm{Kr}}$ &  ${}^{85}{\rm{Rb}}$ &  ${}^{86}{\rm{Sr}}$ &  ${}^{87}{\rm{Sr}}$ &  ${}^{88}{\rm{Sr}}$ &   ${}^{89}{\rm{Y}}$ &  ${}^{90}{\rm{Zr}}$ &  ${}^{92}{\rm{Zr}}$ &  ${}^{93}{\rm{Zr}}$ &  ${}^{94}{\rm{Zr}}$ &\\
$({\rm n},\gamma)$-target &   ${}^{84}{\rm Kr}$ &   ${}^{86}{\rm Kr}$ &   ${}^{87}{\rm Rb}$ &   ${}^{85}{\rm Rb}$ &   ${}^{86}{\rm Sr}$ &   ${}^{87}{\rm Sr}$ &   ${}^{88}{\rm Sr}$ &    ${}^{89}{\rm Y}$ &   ${}^{90}{\rm Zr}$ &   ${}^{92}{\rm Zr}$ &   ${}^{93}{\rm Nb}$ &   ${}^{94}{\rm Zr}$ &\\
$r_{\rm cor,0}$ &$ -0.98$ &$  0.88$ &$  0.86$ &$ -0.86$ &$ -0.94$ &$ -0.92$ &$ -0.65$ &$ -0.83$ &$ -0.88$ &$ -0.92$ &$ -0.97$ &$ -0.85$ &\\
\hline
Product &  ${}^{96}{\rm{Mo}}$ &  ${}^{97}{\rm{Mo}}$ &  ${}^{98}{\rm{Mo}}$ &  ${}^{99}{\rm{Tc}}$ & ${}^{100}{\rm{Ru}}$ & ${}^{102}{\rm{Ru}}$ & ${}^{103}{\rm{Rh}}$ & ${}^{104}{\rm{Pd}}$ & ${}^{106}{\rm{Pd}}$ & ${}^{107}{\rm{Pd}}$ & ${}^{108}{\rm{Pd}}$ & ${}^{109}{\rm{Ag}}$ &\\
$({\rm n},\gamma)$-target &   ${}^{96}{\rm Mo}$ &   ${}^{97}{\rm Mo}$ &   ${}^{98}{\rm Mo}$ &   ${}^{99}{\rm Ru}$ &  ${}^{100}{\rm Ru}$ &  ${}^{102}{\rm Ru}$ &  ${}^{103}{\rm Rh}$ &  ${}^{104}{\rm Pd}$ &  ${}^{106}{\rm Pd}$ &  ${}^{107}{\rm Ag}$ &  ${}^{108}{\rm Pd}$ &  ${}^{109}{\rm Ag}$ &\\
$r_{\rm cor,0}$ &$ -0.94$ &$ -0.87$ &$ -0.94$ &$ -0.91$ &$ -0.92$ &$ -0.86$ &$ -0.95$ &$ -0.97$ &$ -0.96$ &$ -0.80$ &$ -0.96$ &$ -0.79$ &\\
\hline
Product & ${}^{115}{\rm{In}}$ & ${}^{115}{\rm{In}}$ & ${}^{121}{\rm{Sb}}$ & ${}^{126}{\rm{Te}}$ &  ${}^{127}{\rm{I}}$ & ${}^{132}{\rm{Xe}}$ & ${}^{133}{\rm{Cs}}$ & ${}^{134}{\rm{Ba}}$ & ${}^{136}{\rm{Ba}}$ & ${}^{137}{\rm{Ba}}$ & ${}^{138}{\rm{Ba}}$ & ${}^{139}{\rm{La}}$ &\\
$({\rm n},\gamma)$-target &  ${}^{115}{\rm In}$ &  ${}^{115}{\rm Sn}$ &  ${}^{121}{\rm Sb}$ &  ${}^{126}{\rm Te}$ &   ${}^{127}{\rm I}$ &  ${}^{132}{\rm Xe}$ &  ${}^{133}{\rm Cs}$ &  ${}^{134}{\rm Ba}$ &  ${}^{136}{\rm Ba}$ &  ${}^{137}{\rm Ba}$ &  ${}^{138}{\rm Ba}$ &  ${}^{139}{\rm La}$ &\\
$r_{\rm cor,0}$ &$ -0.97$ &$ -0.65$ &$ -0.92$ &$ -0.68$ &$ -0.92$ &$ -0.97$ &$ -0.89$ &$ -0.85$ &$ -0.88$ &$ -0.84$ &$ -0.65$ &$ -0.88$ &\\
\hline
Product & ${}^{159}{\rm{Tb}}$ & ${}^{165}{\rm{Ho}}$ & ${}^{166}{\rm{Er}}$ & ${}^{167}{\rm{Er}}$ & ${}^{168}{\rm{Er}}$ & ${}^{169}{\rm{Tm}}$ & ${}^{181}{\rm{Ta}}$ & ${}^{187}{\rm{Os}}$ & ${}^{192}{\rm{Pt}}$ & ${}^{194}{\rm{Pt}}$ & ${}^{200}{\rm{Hg}}$ & ${}^{205}{\rm{Pb}}$ &\\
$({\rm n},\gamma)$-target &  ${}^{159}{\rm Tb}$ &  ${}^{165}{\rm Ho}$ &  ${}^{166}{\rm Er}$ &  ${}^{167}{\rm Er}$ &  ${}^{168}{\rm Er}$ &  ${}^{169}{\rm Tm}$ &  ${}^{181}{\rm Ta}$ &  ${}^{187}{\rm Os}$ &  ${}^{192}{\rm Pt}$ &  ${}^{194}{\rm Pt}$ &  ${}^{200}{\rm Hg}$ &  ${}^{205}{\rm Tl}$ &\\
$r_{\rm cor,0}$ &$ -0.80$ &$ -0.68$ &$ -0.81$ &$ -0.78$ &$ -0.86$ &$ -0.90$ &$ -0.84$ &$ -0.86$ &$ -0.89$ &$ -0.90$ &$ -0.67$ &$ -0.87$ &\\
\hline
  \end{tabular}
  }
  \label{tab-key-ms}
\end{table}

\section{Conclusion}

We have evaluated the impact on s-process nucleosynthesis in massive stars and low mass AGB stars of nuclear physics uncertainties using a Monte Carlo driven variational technique. We find that (n,$\gamma$) reactions dominate the total uncertainty, with a few important contributions from $\beta$-decays around branching points.  We have then identified individual key reactions in a rigorous and robust way, to guide and support further investigations in nuclear astrophysics regarding the s-process.

\section*{Acknowledgements}

This project has been financially supported by the ERC (EU-FP7-ERC-2012-St Grant 306901-SHYNE, EU-FP7 Adv Grant GA321263-FISH), the UK STFC (ST/M000958/1), ``ChETEC'' COST Action (CA16117) and MEXT as ``Priority Issue on Post-K computer'' (Elucidation of the Fundamental Laws and Evolution of the Universe) and JICFuS. Parts of computations presented in the paper were carried out by computer facilities at the University of Cambridge (COSMOS at DAMPT, STFC DiRAC Facility), University of Edinburgh (Eddie mark 3), CfCA in National Astronomical Observatory of Japan and YITP at Kyoto University.

\bibliography{ref.bib}

\end{document}